\documentclass[conference,9pt]{IEEEtran}
\IEEEoverridecommandlockouts
\usepackage{amsmath,graphicx}
\usepackage{subfigure}
\usepackage{multicol}
\usepackage{multirow}
\usepackage{enumitem} 
\usepackage{booktabs} 
\usepackage{cite}
\usepackage{array}
\usepackage{pifont}
\usepackage{subcaption}
\usepackage{enumitem}
\usepackage{makecell,amsfonts}
\usepackage{pifont}
\usepackage{footnote}
\usepackage{threeparttable}
\usepackage[colorlinks,
            linkcolor=red,
            anchorcolor=blue,
            citecolor=green,
            ]{hyperref}

\def\BibTeX{{\rm B\kern-.05em{\sc i\kern-.025em b}\kern-.08em
    T\kern-.1667em\lower.7ex\hbox{E}\kern-.125emX}}
\begin{document}

\title{UME: Upcycling Mixture-of-Experts for Scalable and Efficient Automatic Speech Recognition}

\author{\IEEEauthorblockN{Li Fu}
\IEEEauthorblockA{\textit{JD AI Research} \\
Beijing, China \\
fuli3@jd.com}
\and
\IEEEauthorblockN{Shanyong Yu}
\IEEEauthorblockA{\textit{JD AI Research} \\
Beijing, China \\
yushanyong@jd.com}
\and
\IEEEauthorblockN{Siqi Li}
\IEEEauthorblockA{\textit{JD AI Research} \\
Beijing, China \\
lisiqi26@jd.com}
\and
\IEEEauthorblockN{Lu Fan}
\IEEEauthorblockA{\textit{JD AI Research} \\
Beijing, China \\
fanlu@jd.com}
\and
\IEEEauthorblockN{Youzheng Wu}
\IEEEauthorblockA{\textit{JD AI Research} \\
Beijing, China \\
wuyouzheng1@jd.com}
\and
\IEEEauthorblockN{Xiaodong He}
\IEEEauthorblockA{\textit{JD AI Research} \\
Beijing, China \\
hexiaodong@jd.com}
}

\maketitle

\begin{abstract}
Recent advancements in scaling up models have significantly improved performance in Automatic Speech Recognition (ASR) tasks. However, training large ASR models from scratch remains costly. To address this issue, we introduce \textbf{UME}, a novel method that efficiently \underline{\textbf{U}}pcycles pretrained dense ASR checkpoints into larger \underline{\textbf{M}}ixture-of-\underline{\textbf{E}}xperts (MoE) architectures. Initially, feed-forward networks are converted into MoE layers. By reusing the pretrained weights, we establish a robust foundation for the expanded model, significantly reducing optimization time. Then, layer freezing and expert balancing strategies are employed to continue training the model, further enhancing performance. Experiments on a mixture of 170k-hour Mandarin and English datasets show that UME: 1) surpasses the pretrained baseline by a margin of 11.9\% relative error rate reduction while maintaining comparable latency; 2) reduces training time by up to 86.7\% and achieves superior accuracy compared to training models of the same size from scratch.
\end{abstract}


\section{Introduction}
\label{sec:intro}
End-to-end architectures have become the predominant approach in Automatic Speech Recognition (ASR) research and applications~\cite{vaswani2017attention,li2022recent,prabhavalkar2023end,zhang2022bigssl,chen2022wavlm,zhang2020pushing}. Scaling up models, such as Whisper~\cite{radford2023robust} and USM~\cite{zhang2023google}, has shown promise in further boosting ASR performance by leveraging larger model capacities and more extensive training data. However, training massive ASR models (e.g., those exceeding 1 billion parameters) from scratch incurs high computational costs and requires intricate hyperparameter tuning and data strategies~\cite{shen2023efficient}. Additionally, simply increasing model size inevitably introduces additional inference latency, posing challenges for real-time applications. Given the inherent generalizability of speech recognition tasks, a well-pretrained smaller model might exhibit comparable capabilities to larger ones in certain respects~\cite{chen2015net2net,wang2023data}. This raises an intriguing question: {\it Can we efficiently scale up models by reusing existing small ASR models as an optimal starting point, thereby reducing training overhead without significantly impacting the Real-Time Factor (RTF)?}

In practice, upcycling pretrained ASR models for scaling up typically involves two main applications:
% \begin{itemize}[noitemsep,topsep=0pt]
% \item Training large ASR models can be hindered by high training costs and tuning complexities. An alternative is to initially train a smaller model and then upcycle it to a larger model once the smaller model saturates~\cite{radford2023robust}.
% \item Enhancing the performance of an existing pretrained ASR model by upscaling it to a larger size.
% \end{itemize}
\begin{itemize}[noitemsep,topsep=0pt]
\item One has access to an existing ASR model and aims to enhance its performance by upscaling to a larger size.
\item One plans to train a large ASR model but is hindered by high training costs and tuning complexities. An alternative is to initially train a small model and then upcycle it to a larger model once the small model saturates~\cite{radford2023robust}.
\end{itemize}

While expanding pretrained ASR models is compelling, it presents challenges including {\it 1) efficient scaling:} developing an upcycling approach that bypasses the high costs of training from scratch and the excessive inference latency of large models, and {\it 2) continued training:} optimizing the training methodology to leverage additional parameters effectively, thus enhancing performance while maintaining existing capability. Mixture-of-Experts (MoE) frameworks have emerged as a pivotal approach for the scalable expansion of neural networks~\cite{artetxe2022efficient,jiang2024mixtral,reid2024gemini,komatsuzaki2022sparse,zhu2024moe}. The core strategy involves replacing the conventional Feed-Forward Networks (FFNs) in dense models with sparsely-gated MoE layers~\cite{shazeer2017outrageously}, which enables model scaling while keeping inference costs low by activating only the top-$k$ expert FFNs~\cite{fedus2022switch}. However, existing MoE-based ASR research typically focuses on training large MoE models from scratch~\cite{you2021speechmoe,you2022speechmoe2,you20223m,wang2023language,chen2023ba,kwon2023mole,hu2023mixture,song2024u2}, a process that can be computationally expensive and require extensive tuning.

\begin{figure}[t]
  \centering
  \includegraphics[width=0.454\textwidth]{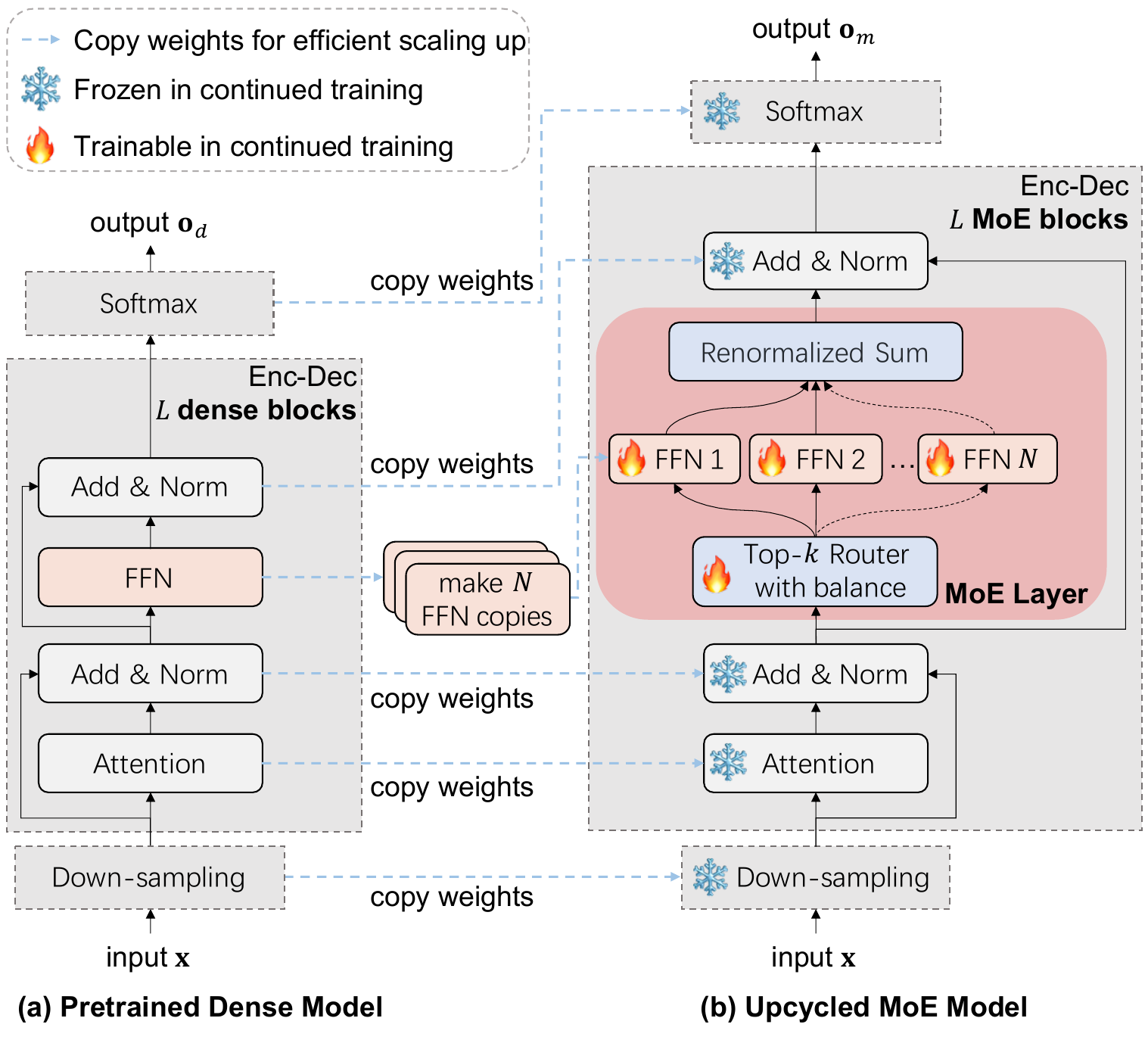}
    \caption{Illustration of the proposed UME. To efficiently scale up, weights from the pretrained dense model are copied to initialize the expanded MoE model. This process ensures an optimal starting point by aligning the MoE model's output ($\mathbf{o}_m$) with that of the dense model ($\mathbf{o}_d$) through a renormalized sum of the top-$\textit{k}$ experts. During continued training, only the MoE layers are unfrozen, with expert balancing, to enhance ASR performance further.}
  \label{fig1}
  \vspace{-0.1in}
\end{figure}
% \footnotetext{Note that the Transformer block~\cite{vaswani2017attention} is used merely as an illustrative example; our method can be easily adapted to other model architectures.}

\begin{table*}[t]
\vspace{-0.2in}
\centering % Center the table

\caption{System performance over training time (days), RTF, and CER/WER for Mandarin/English (M/E) test datasets (\%) (lower is better). Our proposed UME method (C4) consistently outperforms both the pretrained baseline (C0) and Full Model Fine-Tuning (FMFT) (C1) by achieving lower error rates while maintaining a comparable RTF. Additionally, compared to the 1B models trained from scratch (C2 and C3), UME achieves the best performance while significantly reducing training time. ($\star$: Pretraining time could be excluded from the total training cost if one were to have access to the existing model C0.)}

\label{table2} % Label for referencing the table
% Adjust the table size to fit within the page
\resizebox{1.00\linewidth}{!}{
    \begin{tabular}{l|c|c|c|c||cc|c|cccccc}
    \toprule % Horizontal line
    \multirow{2}{*}{\bf Method} & \multirow{2}{*}{\makecell{\bf Model \\ \bf size}} & \multirow{2}{*}{\makecell{\bf Down- \\ \bf sampling}} & \multirow{2}{*}{\makecell{\bf Model \\ \bf type}} & \multirow{2}{*}{\makecell{\bf Training \\ \bf data (h)}} &  \multirow{2}{*}{\makecell{\bf Pretraining \\ \bf time (days)}} & \multirow{2}{*}{\makecell{\bf Continued training \\ \bf time (days)}} & \multirow{2}{*}{\bf RTF} & \multirow{2}{*}{\makecell{\bf SpeechIO \\ \bf (\textit{avg.} 1-26, M)}} & \multirow{2}{*}{\makecell{\bf Wenetspeech \\ \bf test-net (M)}} & \multirow{2}{*}{\makecell{\bf Wenetspeech \\ \bf test-meeting (M)}} & \multirow{2}{*}{\makecell{\bf Aishell-4 \\ \bf meeting (M)}} & \multirow{2}{*}{\makecell{\bf Gigaspeech \\ \bf dev (E)}} & \multirow{2}{*}{\makecell{\bf Gigaspeech \\ \bf test (E)}} \\
     ~ & ~ & ~ & ~ & ~ & ~ & ~ & ~ & ~ & ~ & ~ & ~ & ~ & ~ \\
    \midrule
    C0: Conformer-200M~\cite{gulati2020conformer} & 200M & $\times$8 & Dense & 170k & 5 & NA & \textbf{0.0032} & 3.01 & 6.55 & 6.72 & 16.44 & 13.27 & 13.35  \\
    C1: Conformer-200M-FMFT & 200M & $\times$8 & Dense & 170k & (5$\star$) & 2 & \textbf{0.0032} & 3.00 & 6.56 & 6.70 & 16.48 & 13.29 & 13.32 \\
    \midrule
    C2: Conformer-1B~\cite{gulati2020conformer} & 1B & $\times$8 & Dense & 170k & 15 & NA & 0.0084 & 2.94 & 6.15 & 6.04 & 15.48 & 12.35 & 12.47 \\
    C3: Conformer-1B-MoE~\cite{song2024u2} & 1B & $\times$8 & MoE & 170k & 15 & NA & 0.0042 & 2.92 & 6.09 & 5.99 & 15.35 & 12.20 & 12.31 \\
    \midrule
    C4: Conformer-1B-UME (Ours) & 1B & $\times$8 & MoE & 170k & (5$\star$) & 2 & 0.0042 & \textbf{2.92} & \textbf{6.02} & \textbf{5.92} & \textbf{15.28} & \textbf{12.06} & \textbf{12.13} \\
    \ \ \ \ \ \ \ \ \ \ \ \ \ w/o layer freezing & 1B & $\times$8 & MoE & 170k & (5$\star$) & 3 & 0.0042 & 2.94 & 6.10 & 6.00 & 15.40 & 12.25 & 12.33 \\
    \bottomrule
    \end{tabular}
}
\vspace{-0.1in}
\end{table*}

To efficiently scale up ASR models, we introduce UME, which \textbf{\underline{U}}pcycles pretrained dense checkpoints into larger \textbf{\underline{M}}ixture-of-\textbf{\underline{E}}xperts models to enhance performance. As shown in Fig.~\ref{fig1}, inspired by the advantages of sparsely-gated MoE~\cite{shazeer2017outrageously}, we transform the FFNs of the dense ASR model into MoE layers, thereby expanding the parameter space without significantly increasing inference overhead. To curtail the optimization duration, we repurpose the pretrained checkpoints to initialize the MoE model. This ensures the MoE model aligns with the pretrained model's capabilities, providing a strong starting point for further improvements. During the continued training phase, layer freezing and expert balancing~\cite{fedus2022switch} strategies are adopted to train the additional parameters efficiently. In particular, the training focuses on the MoE experts and the top-\textit{k} router for expert selection, while the remaining parameters are frozen. Besides further reducing training costs, the strategies are designed to alleviate catastrophic forgetting~\cite{li2022massively,fu2021incremental} and boost parameter utilization, resulting in lower error rates. To validate the practicality of our approach, we upcycled state-of-the-art (SOTA) 200M-parameter ASR models~\cite{gulati2020conformer} into 1B-parameter models using large-scale training datasets. Extensive experiments demonstrate that UME significantly outperforms the baselines by achieving substantial accuracy improvement, while also reducing training costs compared to training models of the same size from scratch. Our main contributions are summarized as follows:

1) To the best of our knowledge, this is the first work to efficiently upcycle pretrained dense ASR models into MoE architectures, resulting in enhanced performance.

2) We propose a scaling-up framework called UME, which incorporates weight reusing, layer freezing, and expert balancing strategies to reduce training time and enhance the performance of ASR systems.

3) We demonstrate the effectiveness of UME through diverse datasets and extensive ablation studies, showing that our method significantly reduces error rates compared to strong baselines while maintaining efficiency in both training and inference.

% \vspace{-0.05in}
\section{Related Work}
\label{sec:related}
% \vspace{-0.05in}

\noindent \textbf{MoE for ASR.} The exploration of MoE-based ASR has led to significant advancements. For instance, research on shared embedding networks has improved expert routing mechanisms~\cite{you2021speechmoe, you2022speechmoe2, you20223m}. To enhance multilingual ASR performance, language-based routing has been further examined~\cite{wang2023language, chen2023ba, kwon2023mole}. To simplify model architecture and improve scalability, Hu et al.~\cite{hu2023mixture} introduced a Conformer MoE model for multilingual ASR, notably without using shared embeddings. More recently, Song et al.~\cite{song2024u2} presented a unified MoE model that integrates streaming and non-streaming capabilities, achieving consistent latency levels when scaling a 200M-dense model to a 1B-MoE variant. However, the primary focus of existing research has been developing novel MoE architectures, which often incur high costs when training models from scratch. In contrast, our proposed UME approach emphasizes efficient scaling by upcycling pretrained, smaller dense checkpoints into larger MoE models.

\noindent \textbf{Parameter-efficient fine-tuning for ASR.} In the ASR domain, parameter-efficient fine-tuning methods are explored to add small modules to pretrained models, thereby reducing the computational cost of full-scale training~\cite{sim2024comparison}. For instance, low-rank adapters~\cite{hu2021lora} enable iterative improvements of ASR models by introducing trainable low-rank matrices. However, this approach has primarily been discussed for large pretrained models and may be less effective than full-scale fine-tuning~\cite{sim2024comparison, liu2024sparsely}. While residual adapters~\cite{tomanek2021residual, hou2021exploiting} are also related to enhancing model capacity, they can incur considerable computational load as the size of the adapters increases. Conversely, our method maintains comparable inference costs while effectively boosting model capacity through sparsely-gated MoE.

\section{Our Proposed Approach}
\label{sec:approach}
\subsection{Problem formulation}
\label{subsec:formulation}
\noindent {\bf Pretrained ASR model:} $\mathcal{M}_d$ typically comprises three sequentially stacked modules: 1) a down-sampling module, 2) an encoder-decoder (Enc-Dec) module containing $L$ dense blocks based on the Transformer~\cite{vaswani2017attention} or its variants~\cite{gao2023funasr,gulati2020conformer,gao2022paraformer}, and 3) a softmax module that classifies each token into output characters, as illustrated in Fig.~\ref{fig1}. However, $\mathcal{M}_d$ may not achieve the desired ASR performance due to capacity limitations, data distribution issues, and other factors. \\
\noindent {\bf Training dataset:} $\boldsymbol{D} = \{ (\mathbf{x}^s, \mathbf{y}^s) \mid s \in [1, \ldots, S] \}$, where $S$ is the number of labeled samples. Each $\mathbf{y}^s \in \mathbf{V}^{U_s}$ represents the label sequence (with length $U_s$) of the speech utterance $\mathbf{x}^s$, where $\mathbf{V}$ is the finite set of label characters.\\
\noindent {\bf Aim of UME:} Leverage the dataset $\boldsymbol{D}$ to efficiently scale up the pretrained dense model $\mathcal{M}_d$ into a larger MoE-based model $\mathcal{M}_m$, to enhance accuracy rates without a significant increase in RTF.

\subsection{Model architecture}
\label{subsec:architecture}
\subsubsection{Sparsely-gated MoE structure} 
Building upon the strengths of MoE~\cite{you2021speechmoe,you2022speechmoe2}, we extend the pretrained dense model into an MoE structure by replacing all FFN layers with MoE layers composed of $N$ parallel FFNs. This modification increases the model's parameters, expanding capacity and enhancing performance. To mitigate the computational costs during inference, a routing network is introduced to selectively activate only the top-$k$ experts from the total $N$.

Given input features $\mathbf{h}_{i}\in\mathbb{R}^{T \times d}$ of an MoE layer, which is a sequence of $d$-dimensional features with length $T$, the top-$k$ router is a linear layer that computes normalized weights for expert selection based on $\mathbf{h}_{i}$. Previous works~\cite{you2021speechmoe,you2022speechmoe2} computed the weighted sum of the selected experts' outputs based on the router probabilities over all experts $\mathbf{W}_N = {\rm Softmax}({\rm Linear}(\mathbf{h}_{i}))\in\mathbb{R}^{T \times N}$. However, we found this approach can lead to instability in training when the number of experts varies. To address this issue, we renormalize the weights of the selected top-$k$ expert FFNs $\mathbf{W}_k\in\mathbb{R}^{T \times k}$ using $\mathbf{W}_k = {\rm Softmax}({\rm TopK}({\rm Linear}(\mathbf{h}_{i})))$ and then obtain the output of the MoE layer $\mathbf{h}_{o}\in\mathbb{R}^{T \times d}$ as $\mathbf{h}_{o} = \sum_{n=1}^{k} \mathbf{W}_{k,n} \circ {\rm FFN}_{n}(\mathbf{h}_{i})$, where $\mathbf{W}_{k,n}\in\mathbb{R}^{T \times d}$ is the $n$-th column of the weights $\mathbf{W}_{k}$ repeated $d$ times, and $\circ$ represents element-wise multiplication. This renormalization ensures a robust model initialization, as shown below.

\subsubsection{Pretrained weight reusing}
Training large MoE models from scratch can be costly and unstable~\cite{you2021speechmoe,you2022speechmoe2}. To address the issues, we initialize the MoE model entirely from the pretrained dense checkpoint, substantially reducing the optimization time. Each expert FFN in the MoE layer is copied from the corresponding FFN of the smaller dense checkpoints. Consequently, at the initial stage, the MoE layer and the pretrained FFN layer yield identical results:

\begin{equation}
    \mathbf{h}_{o} = (\sum_{n=1}^{k} \mathbf{W}_{k,n}) \circ {\rm FFN}(\mathbf{h}_{i}) = {\rm FFN}(\mathbf{h}_{i})
    \label{eq2}
\end{equation}

For parameters outside the MoE layers, we directly copy weights from the smaller dense checkpoints. Thus, given input features $\mathbf{x}$, the output of the upcycled MoE model $\mathbf{o}_m$ will be identical to the output of the pretrained dense model $\mathbf{o}_d$. This methodology guarantees that the initial MoE model, irrespective of the number of activated experts, mirrors the recognition performance of the pretrained smaller model. Such alignment provides the expanded MoE model with an optimal starting point, facilitating subsequent enhancements~\cite{wu2024llama}.

\begin{table}[t]
    \vspace{-0.2in}
    \centering
    \caption{System performance over RTF and CER (\%). Our proposed UME method consistently outperforms both the public Paraformer-200M and Full Model Fine-Tuning (FMFT) by achieving lower CERs, without substantially increasing the RTF.}
    \label{table1}
    \resizebox{0.99\linewidth}{!}{
    \begin{tabular}{l|c||c|ccc}
        \toprule
         \multirow{2}{*}{\bf Method}  & \multirow{2}{*}{\makecell{\bf Model \\ \bf type}} & \multirow{2}{*}{\bf RTF} & \multirow{2}{*}{\makecell{\bf Wenetspeech \\ \bf test-net}} & \multirow{2}{*}{\makecell{\bf Wenetspeech \\ \bf test-meeting}} & \multirow{2}{*}{\makecell{\bf Aishell-4 \\ \bf meeting}}  \\ 
         ~  & ~ & ~ & ~ & ~ & ~  \\
        \midrule
        Paraformer-200M~\cite{gao2023funasr} & Dense  & {\bf 0.0061} & 6.74 & 6.97 & 17.63  \\%& \textbf{2.74}
        Paraformer-200M-FMFT  & Dense & {\bf 0.0061} & 6.52 & 6.27 & 15.44  \\ %& 3.56
        Paraformer-1B-UME (Ours)  & MoE & 0.0079 & \textbf{6.30} & \textbf{5.88} & \textbf{15.03} \\ %& 3.09 \\
        \bottomrule
    \end{tabular}
    }
\vspace{-0.1in}
\end{table}

\subsection{Continued training}
\label{subsec:Continued_training}
\subsubsection{Layer freezing} 
A straightforward training approach would be Full Model Fine-Tuning (FMFT) on the dataset $\boldsymbol{D}$. However, applying FMFT to the MoE model might lead to high computational cost and catastrophic forgetting~\cite{li2022massively}. To tackle this challenge, unlike previous methods that introduce dedicated knowledge distillation regularization~\cite{fu2021incremental}, we adopt a simple yet effective layer freezing strategy during the continued training stage. 

As shown in Fig.~\ref{fig1}, we only train the MoE layers, including the expert FFNs and the top-$k$ router, while keeping other parameters frozen, including the down-sampling, attention, and layer normalization modules. This strategy further lowers the training cost by reducing the number of trainable parameters and only refining the simple FFN experts and linear routers. Moreover, by freezing the key pretrained parameters used for hidden feature extraction~\cite{vaswani2017attention,gulati2020conformer}, the MoE model can enhance its performance by adjusting the newly added parameters while preserving its existing robustness~\cite{tomanek2021residual,hou2021exploiting,li2024smartfrz}. Our method proves to be more effective than FMFT in mitigating the issue of catastrophic forgetting (see Section~\ref{subsec:ablation}).

\subsubsection{Expert balancing}
In scaled MoE models, inefficient parameter utilization often arises, with certain experts being overtrained while remaining most underutilized~\cite{shazeer2017outrageously}. To tackle this issue, we introduce the differentiable load balancing loss~\cite{fedus2022switch} $\mathcal{L}_{\rm B}$ into each MoE layer. The loss function encourages a more balanced distribution of tokens across the experts of each MoE layer, as follows:

\begin{equation}
    \mathcal{L}_{\rm B} = N \cdot \sum_{i=1}^{N} (\mathcal{F}_i \cdot \mathcal{G}_i)
    \label{eq3}
\end{equation}

\noindent where $\mathcal{F}_i = \frac{1}{T} \sum_{t=1}^{T} \mathbf{1}{\{\text{argmax}({\bf W}^{t}_{N}) = i\}}$ is the fraction of tokens dispatched to expert $\text{FFN}_i$, and $\mathcal{G}_i = \frac{1}{T} \sum_{t=1}^{T} {\bf W}^{t}_{N,i}$ is the average routing probability of expert $\text{FFN}_i$, where ${\bf W}^{t}_{N}$ is the $t$-th row of ${\bf W}_{N}$, and ${\bf W}^{t}_{N,i}$ is the $i$-th element of ${\bf W}^{t}_{N}$. 

% \noindent where $\mathcal{F}_i = \frac{1}{T} \sum_{t=1}^{T} \mathbf{1}{\{\text{argmax}({\bf W}^{t}_{N}) = i\}}$ is the fraction of tokens dispatched to expert $\text{FFN}_i$, with ${\bf W}^{t}_{N}$ the $t$-th row of ${\bf W}_{N}$; and $\mathcal{G}_i = \frac{1}{T} \sum_{t=1}^{T} {\bf W}^{t}_{N,i}$ is the average routing probability of expert $\text{FFN}_i$, with ${\bf W}^{t}_{N,i}$ the $i$-th element of ${\bf W}^{t}_{N}$.

Finally, the total training loss function $\mathcal{L}_{\rm total}$ is composed of the expert balancing loss function $\mathcal{L}_{\rm B}$ and the conventional ASR loss function $\mathcal{L}_{\rm ASR}$, resulting in 

\begin{equation}
    \mathcal{L}_{\rm total} = \mathcal{L}_{\rm ASR} + \alpha \mathcal{L}_{\rm B}
    \label{eq6}
\end{equation}

\noindent where the coefficient $\alpha$ balances the terms. Following~\cite{fedus2022switch}, we set $\alpha = 0.01$ to ensure load balancing without overwhelming the primary ASR objective. Note that our method can be readily applied to different models and ASR loss functions, as demonstrated in Section~\ref{subsec:main_exp} and Section~\ref{subsec:extend_exp}.

\begin{figure}[t]
\vspace{-0.4in}     
        \centering
	\subfigure[w/o expert balancing]{
		\begin{minipage}[t]{0.35\linewidth}
			\centering
			\includegraphics[width=\linewidth]{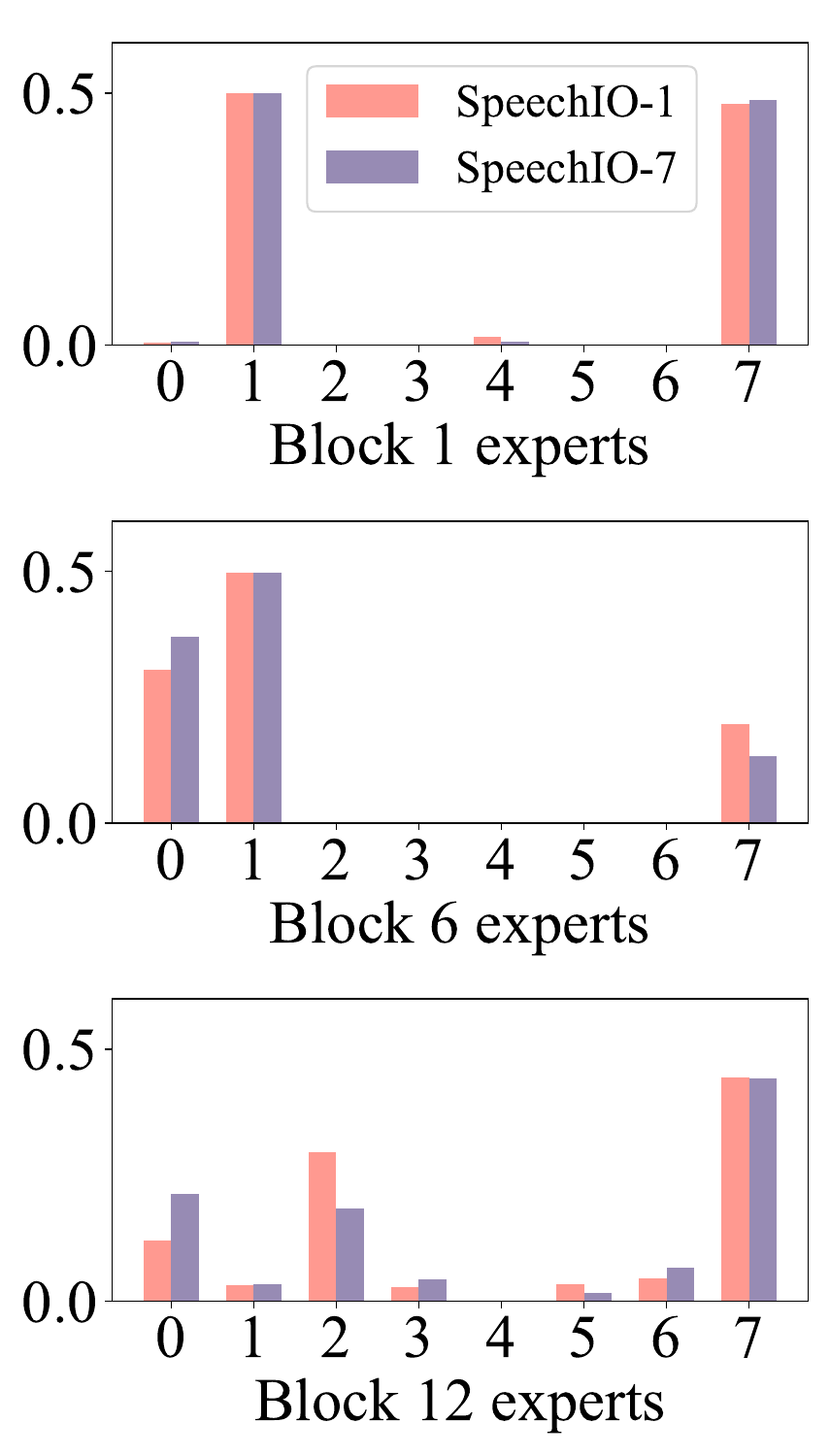}
		\end{minipage}
	}%
        % \vspace{0.1em}
	\subfigure[w/ expert balancing]{
		\begin{minipage}[t]{0.35\linewidth}
			\centering
			\includegraphics[width=\linewidth]{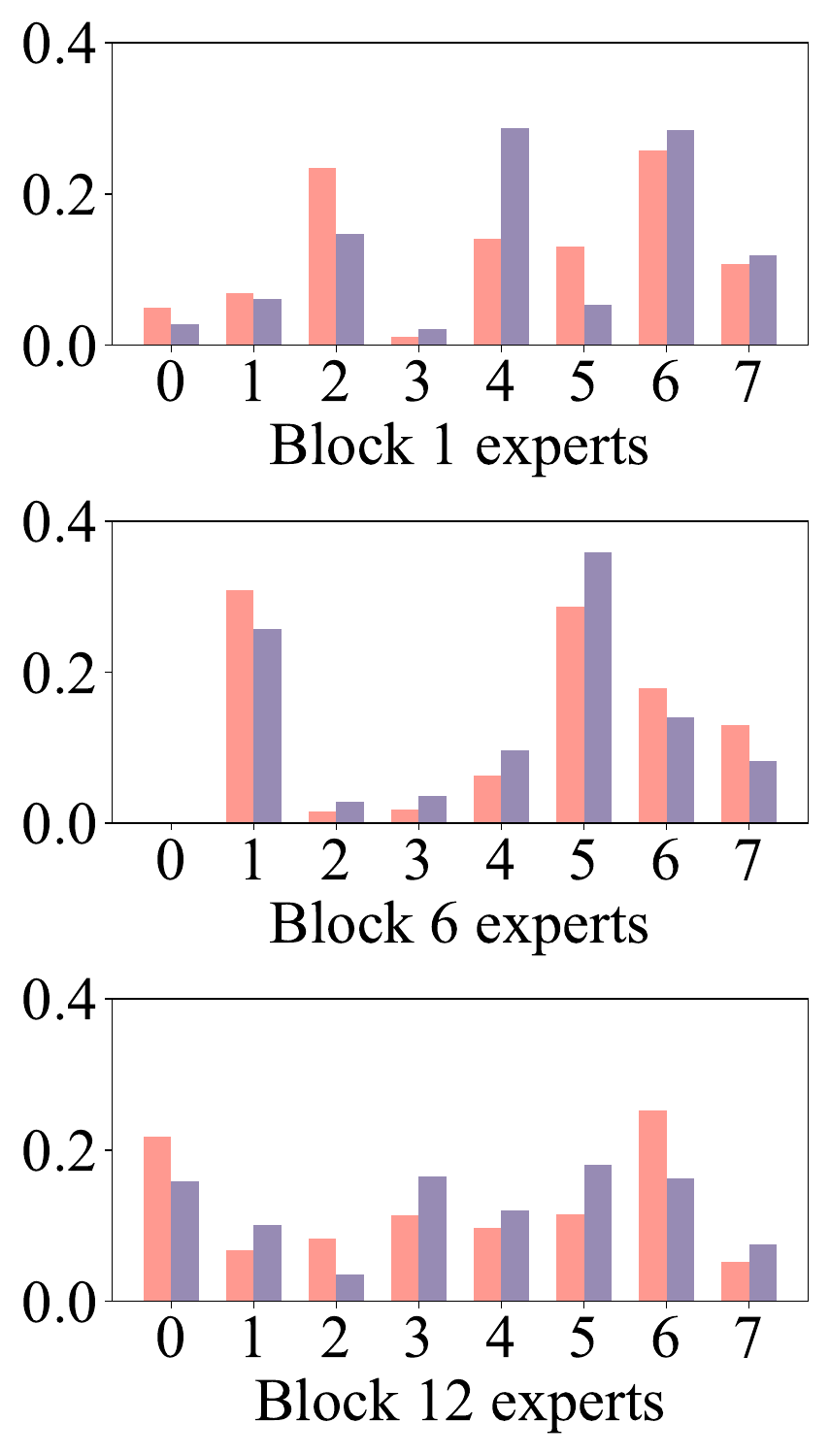}
		\end{minipage}
	}%
	\centering
        \caption{Distribution of activated experts across Conformer-1B-UME blocks with and without expert balancing.}
	\label{fig2}
        \vspace{-0.2in}
\end{figure}

\section{Experiments and Discussion}
\label{sec:experiments}
\subsection{Experimental setup}
\label{subsec:exp_setup}

{\noindent \bf Data preparation.} All speech samples are at a 16kHz sampling rate and are preprocessed into 80-dimensional filter banks using a 25ms window with a 10ms shift.

\noindent \textit{• Training Set:} We conducted experiments using a large-scale dataset to improve model scalability and generalization. This dataset consists of 170k hours of labeled data, primarily in Mandarin (90\%) with the remaining in English (10\%). It covers a wide range of domains, including internet content and recorded meetings.

\noindent  \textit{• Testing Set:} To mitigate bias and uncover distinct challenges, we validated the ASR systems with five public test sets, each derived from internet and meeting scenarios:
\ding{172} the SpeechIO benchmark\footnote{https://github.com/SpeechColab/Leaderboard}, which encompasses 26 varied internet scenarios for a comprehensive assessment;
\ding{173} Wenetspeech test-net~\cite{zhang2022Wenetspeech}, a internet multi-domain test set;
\ding{174} Wenetspeech test-meeting~\cite{zhang2022Wenetspeech} and
\ding{175} Aishell-4 meeting~\cite{fu2021aishell}, both sourced from actual meetings to present more challenging tests; \ding{176} Gigaspeech dev/test~\cite{chen2021gigaspeech}, used to evaluate English performance.

{\noindent \bf ASR models\footnote{Our code and models will be released upon publication.}.} To address scaling-up scenarios, we developed SOTA Conformer-200M models (C0) with 200M parameters. This autoregressive model comprises 12 encoder blocks and 6 bidirectional decoder blocks with an 8-fold down-sampling rate. For more details on the model structure, refer to~\cite{song2024u2}. Initially, we pretrained C0 with the training dataset, and then scaled it to 1B parameters, resulting in Conformer-1B-UME (C4). FMFT was applied to obtain Conformer-200M-FMFT (C1) for comparison. Additionally, a dense Conformer-1B model (C2) and a 1B-MoE model, Conformer-1B-MoE (C3), were trained from scratch to test the efficiency of our method.

{\noindent \bf Training \& Decoding.} All models were trained on 8 NVIDIA A100 GPUs, with each GPU processing 400 seconds of speech per batch. The Adam optimizer was used with a learning rate of 2e-4, which was warmed up over the first 10k steps. Early stopping was applied when the validation loss did not decrease for 6k training steps. Following~\cite{song2024u2}, we employed CTC greedy search decoding~\cite{graves2006connectionist} to achieve efficient inference.

{\noindent \bf Performance evaluation.} We directly calculated the CER/WER on the Mandarin/English test sets to evaluate the models' recognition performance. For clarity, we averaged the CERs across the 26 scenarios from the SpeechIO benchmark to assess general performance. To measure the RTF, we set the batch size to 20 and performed inference using a single NVIDIA Tesla V100 GPU.

\begin{table}[t]
\vspace{-0.2in}
\centering % Center the table
% \caption{CER results with and without expert balancing.}
\caption{Ablation studies on layer freezing and expert balancing strategies in CER ($\%$).}
\label{table3} % Label for referencing the table
\resizebox{0.98\linewidth}{!}
{
    \begin{tabular}{l||c|c|c|c}
    \toprule
    % \multirow{2}{*}{\bf Method} & \multirow{2}{*}{\makecell{\bf SpeechIO \\ \bf (\textit{avg.} 1-26)}}  & \multirow{2}{*}{\makecell{\bf Wenetspeech \\ \bf test-net}} & \multirow{2}{*}{\makecell{\bf Wenetspeech \\ \bf test-meeting}} & \multirow{2}{*}{\makecell{\bf Aishell-4 \\ \bf meeting}} \\
    \multirow{2}{*}{\bf Method} & \multirow{2}{*}{\makecell{\bf SpeechIO \\ \bf (\textit{avg.} 1-26)}}  & \multirow{2}{*}{\makecell{\bf Wenetspeech \\ \bf test-net}} & \multirow{2}{*}{\makecell{\bf Wenetspeech \\ \bf test-meeting}} & \multirow{2}{*}{\makecell{\bf Aishell-4 \\ \bf meeting}} \\
    ~ & ~ & ~ & ~ & ~ \\
    \midrule
    Conformer-1B-UME&  \textbf{2.92} &\textbf{6.02} & \textbf{5.92} & \textbf{15.28}  \\
    \ \ \ \ \ w/o layer freezing & 2.94 & 6.10 & 6.00 & 15.40  \\
    \ \ \ \ \ w/o expert balancing & 2.96 & 6.21 & 6.10 & 15.54  \\
    \midrule
    Paraformer-1B-UME & \textbf{3.09} & \textbf{6.30} & \textbf{5.88} & \textbf{15.03}  \\
    \ \ \ \ \ w/o layer freezing & 3.32 & 6.40 & 6.01 & 15.22  \\
    \ \ \ \ \ w/o expert balancing & 3.18 & 6.41 & 6.17 & 15.58  \\
    % \ \ \ \ \ \ w/ expert balancing &  \\
    \bottomrule
    \end{tabular}
}
\vspace{-0.1in}
\end{table}

\subsection{Main experiments: Conformer with 170k-hour dataset}
\label{subsec:main_exp}

{\noindent \bf Enhancement in recognition performance.} As shown in Table~\ref{table2}, our UME-based model C4 consistently achieves CER reductions across all test sets compared to the pretrained dense baseline model C0 and the FMFT method C1. Numerically, UME achieves a maximum relative CER reduction of 11.9\% compared to C0. Note that this improvement is achieved without a significant increase in RTF, benefiting from the sparsely-gated MoE structure. Despite continued training on the same data, the performance of FMFT does not improve compared to C0. We infer that the model reaches saturation on the large-scale dataset. In contrast, our UME method substantially enhances performance by increasing model capacity.

{\noindent \bf Efficiency in training time.} Table~\ref{table2} demonstrates that the UME method largely reduces training cost compared to the 1B models trained from scratch (C2 and C3) while also delivering superior CER performance. Numerically, by starting with the pretrained model C0, the expanded model reduces 86.7\% time compared to training the 1B models from randomly initialized weights. This approach is particularly beneficial when one has access to an existing ASR model and aims to enhance its performance by scaling up to a larger size. In particular, the layer freezing strategy further reduced the required training time by 33.3\% by decreasing the number of trainable parameters. Even when considering the time needed to pretrain the model, our method still reduces overall computational costs by 53.3\%. This demonstrates that leveraging pretrained parameters provides an optimal starting point, facilitating subsequent enhancements while significantly reducing optimization time.

\subsection{Extended experiment: Paraformer with public 10k-hour dataset} 
\label{subsec:extend_exp}
To further validate the effectiveness of our approach and ensure reproducibility in research, we conducted extended experiments on different ASR model structures and varying sizes of training datasets. Specifically, we upcycled the Paraformer-200M model~\cite{gao2023funasr} using a public 10k-hour dataset, which combines Wenetspeech~\cite{zhang2022Wenetspeech} and Aishell-4~\cite{fu2021aishell}. Paraformer-200M is an open-source, non-autoregressive Mandarin ASR model with 200 million parameters. It comprises 50 encoder blocks and 16 decoder blocks, featuring a 6-fold down-sampling rate~\cite{gao2022paraformer}. Notably, the open-source model was pretrained on 60k hours of data~\cite{gao2023funasr}. Our results align with those observed for the Conformer models. As shown in Table~\ref{table1}, the proposed UME method (Paraformer-1B-UME) achieves the lowest CER values on all test sets, with a maximum relative CER reduction of 15.6\% compared to the pretraining baseline (Paraformer-200M), without substantially increasing the RTF. Compared to the FMFT approach (Paraformer-200M-FMFT), our proposed method consistently achieves superior CER values by converting dense FFNs into high-capacity MoE layers.

\subsection{Ablation study and discussion}
\label{subsec:ablation}
To investigate the impact of different components in UME, we conducted the following ablation studies and discussions.

{\noindent {\bf Layer freezing for mitigating forgetting.}} To analyze the effects of the layer freezing strategy during continued training, we conducted ablation experiments on both the Conformer and Paraformer models. As shown in Table~\ref{table3}, incorporating layer freezing consistently decreases CER across all test sets. Notably, compared to the pretrained Paraformer-200M, which achieves a 2.74\% CER on the SpeechIO benchmark, both FMFT (with a 3.56\% CER) and UME without layer freezing (with a 3.32\% CER) suffer from significant forgetting issues. We infer that the difference in distribution between the continued training data and the SpeechIO benchmark leads to this occurrence. By employing the layer freezing strategy, our approach can markedly mitigate forgetting, resulting in a relative CER reduction of 6.9\% on the SpeechIO benchmark compared to not using this strategy. Our results demonstrate that freezing pretrained parameters, such as those in attention modules, helps preserve the model's existing capability and enhances performance by adjusting newly added parameters.

{\noindent {\bf Expert balancing for boosting expert usage.} Contrary to existing work~\cite{song2024u2} that omitted expert balancing to simplify the training process, we discovered that expert balancing consistently reduces CER across all test sets compared to not using it, as shown in Table~\ref{table3}. Moreover, we found that the balancing strategy improves the expert usage across different scenarios (see Fig.~\ref{fig2}). To investigate this effect, we analyzed expert usage on two test sets: SpeechIO-1 (a news broadcast scenario with stable speech rates and factual content) and SpeechIO-7 (a live streaming sales scenario with variable speech rates and product introductions). We calculated the distribution of activated experts across selected blocks of the Conformer-1B-UME model for the two distinct scenarios. With balancing, more experts are engaged, and a clear distinction in expert usage between the two scenarios emerges. We infer that expert balancing enables the model to leverage parameters more efficiently, resulting in lower CERs.

{\noindent {\bf Tuning expert numbers and top-\textit{k} selections.} We investigated the impact of varying the number of experts and the top-\textit{k} selection in the MoE layers. As shown in Table~\ref{table4}, increasing the number of experts from 4 to 8 and using a top-\textit{k} of 2 consistently improves performance. This increase in both hyperparameters enhances model capacity and boosts CER performance. Importantly, since only the top-\textit{k} experts are selectively activated during inference, our proposed UME method does not incur excessive computational costs. Our results are promising and suggest a potential avenue for further performance gains by increasing the hyperparameters beyond the current settings. However, such enhancements should be carefully weighed against the corresponding increase in computational demand. In our study, we opted for a configuration of 8 experts and a top-\textit{k} of 2 as a balanced choice that offers significant performance improvements without incurring prohibitive computational costs. Future research will explore larger configurations to enhance performance further.

\begin{table}[t]
\vspace{-0.2in}
\centering % Center the table
\caption{Performance of UME for Conformer with varying numbers of experts and top-\textit{k} selections in RTF and CER/WER (\%).}
\label{table4} % Label for referencing the table
% Adjust the table size to fit within the page
\resizebox{0.98\linewidth}{!}
{
    \begin{tabular}{c|c||c|c|c|c|c}
    \toprule % Horizontal line
    % \multirow{2}{*}{\makecell{\bf Number \\ \bf of experts}} & \multirow{2}{*}{\makecell{\bf Top-\textit{k} \\ \bf selection}} &\multirow{2}{*}{\bf RTF}  & \multirow{2}{*}{\makecell{\bf SpeechIO \\ \bf (\textit{avg.} 1-26)}} & \multirow{2}{*}{\makecell{\bf Wenetspeech \\ \bf test-net}} & \multirow{2}{*}{\makecell{\bf Wenetspeech \\ \bf test-meeting}} & \multirow{2}{*}{\makecell{\bf Aishell-4 \\ \bf meeting}}  \\
    % {\bf Number of experts} & {\bf Top-\textit{k} selection} &{\bf RTF}  & {\bf SpeechIO (\textit{avg.} 1-26)} & {\bf Wenetspeech test-net} & {\bf Wenetspeech test-meeting} & {\bf Aishell-4  meeting}  \\
    % % ~ & ~ & ~ & ~ & ~ & ~ \\
    \multirow{2}{*}{\makecell{\bf Number \\ \bf of experts}} & \multirow{2}{*}{\makecell{\bf Top-\textit{k} \\ \bf selection}} &\multirow{2}{*}{\bf RTF}  & \multirow{2}{*}{\makecell{\bf SpeechIO \\ \bf (\textit{avg.} 1-26)}} & \multirow{2}{*}{\makecell{\bf Wenetspeech \\ \bf test-net}} & \multirow{2}{*}{\makecell{\bf Wenetspeech \\ \bf test-meeting}} & \multirow{2}{*}{\makecell{\bf Gigaspeech \\ \bf test}}  \\
    ~ & ~ & ~ & ~ & ~ & ~ \\
    \midrule
    $N$=4 & $k$=1 & {\bf 0.0034} & 3.00 & 6.37 & 6.46 & 12.98  \\

    $N$=4 & $k$=2 & 0.0041 & 2.97 &6.24 & 6.33 & 12.71  \\

    $N$=8 & $k$=1 & 0.0035 & 2.94 & 6.19 & 6.02 & 12.41  \\

    $N$=8 & $k$=2 & 0.0042 & \textbf{2.92} & \textbf{6.02} & \textbf{5.92} & \textbf{12.13}  \\
    \bottomrule
    \end{tabular}
}
\vspace{-0.1in}
\end{table}

\section{Conclusions}
\label{sec:conclusion_futurework}
In this work, we introduced UME, a novel method for upcycling pretrained dense ASR models into larger MoE models, thereby enabling efficient scaling. By reusing pretrained weights, we establish a robust starting point, significantly reducing optimization time. The model is further enhanced through the incorporation of layer freezing and expert balancing techniques. Extensive experiments demonstrate that UME consistently outperforms baseline models, achieving substantial reductions in both error rates and training costs.

\bibliographystyle{IEEEbib}
\bibliography{strings}
\end{document}